\documentclass[letterpaper, 10 pt, conference]{ieeeconf}  
\pdfoutput=1

\IEEEoverridecommandlockouts                              

\usepackage{tikz}
\usetikzlibrary{arrows, shapes, positioning, shapes.geometric, decorations.markings, calc}
\usepackage{siunitx}
\usepackage{tabularx}
\usepackage{booktabs}
\usepackage[caption=false, font=footnotesize]{subfig}
\usepackage{cite}

\usepackage{hologo}
\usepackage{listings}
\title{\LARGE \bf
Towards Efficient Hazard Identification in the Concept Phase \\of Driverless Vehicle Development*
}

\author{Robert Graubohm$^{1}$, Torben Stolte$^{1}$, Gerrit Bagschik$^{2}$, and Markus Maurer$^{1}$
\thanks{*This research is accomplished within the project ``UNICAR\emph{agil}'' (FKZ~16EMO0285). We acknowledge the financial support for the project by the Federal Ministry of Education and Research of Germany (BMBF).}
\thanks{$^{1}$Robert Graubohm, Torben Stolte and Markus Maurer are with the Institute of Control Engineering at Technische Universit\"at Braunschweig, 38106 Braunschweig, Germany
        {\tt\small \{graubohm,stolte,maurer\}@ifr.ing.tu-bs.de}}%
\thanks{$^{2}$Gerrit Bagschik is currently employed at Zoox, Inc. Contributions to this paper were performed during employment at the Institute of Control Engineering at Technische Universit\"at Braunschweig}%
}

\begin{document}

\twocolumn[
\begin{@twocolumnfalse}
			\Huge {IEEE copyright notice} \\ \\
	\large {\copyright\ 2020 IEEE. Personal use of this material is permitted. Permission from IEEE must be obtained for all other uses, in any current or future media, including reprinting/republishing this material for advertising or promotional purposes, creating new collective works, for resale or redistribution to servers or lists, or reuse of any copyrighted component of this work in other works.} \\ \\
	
	{\Large Published in \emph{2020 IEEE Intelligent Vehicles Symposium (IV)}, Las Vegas, NV, USA, October 19--November 13, 2020.} \\ \\ 
	
	{\Large DOI: \href{https://doi.org/10.1109/IV47402.2020.9304780}{10.1109/IV47402.2020.9304780}} \\ \\

Cite as:
\vspace{0.1cm}

\noindent\fbox{%
    \parbox{\textwidth}{%
        R.~Graubohm, T.~Stolte, G.~Bagschik, and M.~Maurer, ``Towards {Efficient} {Hazard} {Identification} in the {Concept} {Phase} of {Driverless} {Vehicle} {Development},''
  in \emph{2020 {IEEE} {Intelligent} {Vehicles} {Symp.}},\hskip 1em plus
  0.5em minus 0.4em\relax Las~Vegas, NV, USA, 2020, pp. 1297--1304, doi:
  {10.1109/IV47402.2020.9304780}.
    }%
}
\vspace{2cm}

\end{@twocolumnfalse}
]

\noindent\begin{minipage}{\textwidth}

\hologo{BibTeX}:
\footnotesize
\begin{lstlisting}[frame=single]
@inproceedings{graubohm_towards_2020,
  author={{Graubohm}, Robert and {Stolte}, Torben and {Bagschik}, Gerrit and {Maurer}, Markus},
  booktitle={2020 IEEE Intelligent Vehicles Symposium (IV)},
  title={Towards {Efficient} {Hazard} {Identification} in the {Concept} {Phase} of {Driverless} {Vehicle } {Development}},
  address={Las Vegas, NV, USA},
  year={2020},
  pages={1297--1304},
  doi={10.1109/IV47402.2020.9304780},
  publisher={IEEE}
}
\end{lstlisting}
\end{minipage}

\maketitle
\thispagestyle{empty}
\pagestyle{empty}

\begin{abstract}

The complex functional structure of driverless vehicles induces a multitude of potential malfunctions. Established approaches for a systematic hazard identification generate individual potentially hazardous scenarios for each identified malfunction. This leads to inefficiencies in a purely expert-based hazard analysis process, as each of the many scenarios has to be examined individually. In this contribution, we propose an adaptation of the strategy for hazard identification for the development of automated vehicles. Instead of focusing on malfunctions, we base our process on deviations from desired vehicle behavior in selected operational scenarios analyzed in the concept phase. By evaluating externally observable deviations from a desired behavior, we encapsulate individual malfunctions and reduce the amount of generated potentially hazardous scenarios. After introducing our hazard identification strategy, we illustrate its application on one of the operational scenarios used in the research project UNICAR\emph{agil}. 

\end{abstract}

\section{INTRODUCTION}

\label{sec:introduction}

Conducting a hazard analysis poses a major challenge within the development of safety-critical automated driving functions.
The identification of unacceptable risks due to hazards is essential for a safety concept development.
In order to identify hazards, malfunctioning behaviors of the vehicle under design are considered in context of its operational scenarios.
To describe resulting scenarios where the vehicle poses hazards to health and life of humans, we propose the use of the term \emph{hazardous scenario}~\cite{stolte_hazard_2017}.

Instructions for the systematic identification of hazardous scenarios can be found in the functional safety standard ISO~26262~\cite[Part~3,~6.4.2]{international_organization_for_standardization_iso_2018}.
The application of these instructions in the development of driverless vehicles, however, remains an open problem.
In contrast to conventional vehicle electronics, the system boundaries and operational context of driverless vehicles for public roads result in a vast amount of possible operational scenarios and potential malfunctions.
Bagschik et al.~\cite{bagschik_identification_2016} demonstrate this problem:
closely following the instructions of ISO~26262, they generate more than 21,000 potentially hazardous scenarios for a limited use case of driverless operation.

Finding hazardous scenarios and their functional causes already in early design stages is necessary to iteratively develop safety mechanisms and to define appropriate functional system boundaries from the beginning.
Additionally, with regard to SAE level 3+ vehicles~\cite{sae_international_sae_2018}, safety aspects to be analyzed exceed the functional safety perspective~\cite{feth_multi-aspect_2018,bagschik_systems_2018}.
Therefore, an approach to systematic hazard identification for driverless vehicles has to enable hazard analyses in early design stages and from multiple safety perspectives.

Prior to the hazard identification, the safety life cycle of ISO~26262 requires to specify the functionalities, boundaries, operational conditions, etc. of the system under development in an \emph{item definition}.
In order to include safety of the intended functionality (SOTIF)~\cite{international_organization_for_standardization_iso/pas_2019} and behavioral safety~\cite{waymo_waymo_2017} considerations, we develop safety concepts on a vehicle level in the concept phase.
Consequently, when generating the item definition, multiple systems and mechatronic components of a vehicle are combined within one item.
On the one hand, the macroscopic description of vehicle functionalities supports the definition of hazards on the vehicle level, as required by ISO~26262.
On the other hand, many functions bringing forth multiple possible malfunctions have to be analyzed in a single hazard identification process.
Evidently, the combination of all potential malfunctions with all operational scenarios is not feasible for complex automated driving systems.

Currently, we are in the process of developing a safety concept for driverless vehicles in inner city traffic as one of the goals in the research project UNICAR\emph{agil}, which is further introduced in Section~\ref{sec:usecase}.
In order to account for safety aspects beyond functional safety and to enable a hazard analysis and risk assessment in the large operational design domain, we adapted the conventional hazard identification strategy and discuss our alterations in this contribution.
First, the hazard identification in related work is discussed in the following section.
Subsequently, Section~\ref{sec:proposedstrat} introduces our proposed adaptation of the identification of potential malfunctioning behaviors.
To illustrate the manageable amount of generated potentially hazardous scenarios, Section~\ref{sec:usecase} presents an application of our approach in the research project UNICAR\emph{agil}.
Lastly, Section~\ref{sec:comparison} gives a comparison of the proposed strategy for hazard identification with other published processes.

\section{PUBLISHED HAZARD IDENTIFICATION STRATEGIES}
\label{sec:strategies}

\subsection{Hazard Identification Instructions of ISO~26262}

\label{ssec:isostrategy}

Within the development process of ISO~26262, the identification of hazardous scenarios caused by malfunctioning behavior of the item is part of the hazard analysis and risk assessment task. The situation analysis performed for hazard identification shall combine operational situations, operating modes, and an item's malfunctioning behavior~\cite[Part~3,~6.4.2.1]{international_organization_for_standardization_iso_2018}. Operational situations represent intended use and foreseeable misuse of the system under development within its operational design domain. Menzel et al.~\cite{menzel_scenarios_2018} propose the use of the term \emph{operational scenario} instead of operational situation in context of the definition by Ulbrich et al.~\cite{ulbrich_defining_2015}, which has been picked up by the ISO/PAS~21448 (SOTIF) standard~\cite{international_organization_for_standardization_iso/pas_2019}.

ISO~26262 demands a systematical determination of hazards based on possible malfunctioning behaviors of an item, which is described as the loss of one or several functions. The standard suggests the use of failure mode and effects analysis (FMEA) approaches and a hazard and operability study (HAZOP) to support the investigation. The effects of the malfunctioning behavior of an item are analyzed subsequently to identify hazards. In order to allow a risk classification by evaluation of potential consequences of each hazard, ISO~26262 requires hazards to be defined at the vehicle level. A distinct feature of the hazard identification is the presumption of correct functionality of every other sufficiently independent system when identifying hazards caused by malfunctioning behavior of the item~\cite[Part~3,~6.4.2.3]{international_organization_for_standardization_iso_2018}.

The hazard identification strategy within the development process of ISO~26262 is widely adopted by concepts and case studies described in related work (cf.~Section~\ref{ssec:relatedwork}). 
However, Van Eikema Hommes~\cite{van_eikema_hommes_review_2012} points out an overall lack of specific guidelines in ISO~26262 on how to conduct hazard identification as a structured process.

\subsection{Related Work}

\label{ssec:relatedwork}

Bagschik et al.~\cite{bagschik_identification_2016} describe the process of systematic identification of hazardous scenarios on a vehicle level in the context of the development of an unmanned protective vehicle for hard shoulder roadworks.
Their work includes a discussion on the scientific and normative background of the term \emph{hazard} in the context of automated driving.
Due to the limited applicability of traditional hazard analysis techniques in the concept phase of development, they utilize skill graphs presented by Reschka et al.~\cite{reschka_ability_2015} as a functional model of the system to identify potential system malfunctions.
In context of the same project, Stolte et al.~\cite{stolte_hazard_2017} present the process and results of an expert-based hazard analysis and risk assessment for the vehicle guidance system.

Based on research by B\"uker et al.~\cite{buker_identifikation_2019}, the research project PEGASUS uses a keyword-based strategy to identify hazards~\cite{pegasus_project_office_pegasus_nodate}.
To find hazardous scenarios, they analyze incorrect vehicle behavior due to component failure in the context of operational scenarios.
Subsequently, uncovered potential hazards are an input for the identification of automation risks of the introduction of a highly automated driving function into public traffic.

Valdez Banda et al.~\cite{valdez_banda_systemic_2019} apply a systematic hazard analysis for early design stages to autonomous vessel concepts for urban transport.
They identify 15 hazards that can lead to accidents by analyzing system states and operational conditions.
The definition of ten potential accidents is based on statistical data and experts' judgement.
In a subsequent process step, Valdez Banda et al. detail their hazard descriptions by incorporating causal factors, potential severity, and types of consequences.

Aceituna~\cite{aceituna_prehaz:_2019} performs a hazard assessment for an autonomous vehicle in an urban environment.
He uses domain models to narrow the hazard space analyzed in an expert-based hazard identification.
Aceituna details possible causes and safety measures for 19 hazards in the interaction of an autonomous vehicle with its environment.

The following publications (i.e.~\cite{alexander_deriving_2009,lurie_hazard_2018,arai_scenario-based_2019,dreany_cognitive_2019}) discuss the identification of hazards as part of a safety concept development, but do not indicate a systematic process to find relevant malfunctioning behaviors.
For example, Alexander et al.~\cite{alexander_deriving_2009} introduce a hazard identification based on an energy trace and barrier analysis as well as checklists as a first step of safety requirement analysis for an unmanned aerial vehicle.
Sch\"onemann et al.~\cite{arai_scenario-based_2019} discuss the process and results of a systematic hazard analysis and risk assessment process for an automated valet parking system.
They propose to perform the hazard identification in each individual scenario based on potential malfunctions of the system.
To address multiple safety aspects and analyze potential accidents, Lurie and Miller~\cite{lurie_hazard_2018} propose a Markov process that describes transitions between driving situations and hazards.
Based on Kurd et al.~\cite{kurd_developing_2006}, Dreany and Roncace~\cite{dreany_cognitive_2019} identify hazards, causal factors, and generic mishap by iteratively assessing the application and functional structure of an unmanned surface vehicle.

Other publications discuss the hazard identification process in the context of the development of vehicle subsystems in conventional vehicles.
For example, Beckers et al.~\cite{beckers_structured_2013} apply a model-based hazard analysis method to an electronic steering column lock.
They use guidewords to identify potential failures based on the specified functions of the item, and subsequently identify hazards by analyzing the effect of malfunctions observed at the vehicle level in a set of scenarios.
Similarly, Becker et al.~\cite{becker_functional_2018} combine malfunctioning behaviors of an electric power steering system with operational scenarios to determine hazards.
Sexton et al.~\cite{sexton_effective_2014} perform a systematic identification of hazards for a shift-by-wire system.
Moreover, an experimental study of two methods for hazard identification in the early phases of development of an electronic steering column lock system has been performed by T\"orner et al.~\cite{hutchison_assessment_2006}.
They find the induction of generic failure modes on a system level to be more efficient than the application of generic low-level hazards to components.

Altogether, most published approaches for a systematic hazard identification are limited to functional safety. In contrast, safety analyses beyond functional safety considerations require a wider perspective towards sources of potential malfunctioning behavior. ISO/PAS~21448~\cite{international_organization_for_standardization_iso/pas_2019}, for example, specifically addresses hazards due to performance limitations and foreseeable misuse. Both types of hazards remain unidentified when strictly and solely performing hazard identification for functional safety. The SOTIF standard, however, adopts the overall hazard identification strategy of ISO~26262 by analyzing unintended behavior caused by triggering events within specific scenarios~\cite[Clause~6.2]{international_organization_for_standardization_iso/pas_2019}.

The hazard identification performed in related work is widely based on combining operational scenarios with all identified potential system failures.
Here, the presumably high number of possible malfunctions of driverless vehicles results in an amount of potentially hazardous scenarios that can likely hinder an efficient expert-based hazard analysis.
In contrast, a manageable amount of generated potentially hazardous scenarios is required for the integration of a systematic identification approach in the safety analysis process of the concept phase.
The adaptation of the identification strategy of potential malfunctioning behaviors proposed in the following section aims to address this problem.

\section{PROPOSED HAZARD IDENTIFICATION PROCEDURE}

\label{sec:proposedstrat}

The implementation of a hazard identification process strongly relies on the definition of possible malfunctioning behaviors before putting the vehicle behavior in context of operational scenarios. The strategies for hazard identification described in Section~\ref{ssec:isostrategy} infer potential malfunctioning behaviors from potential system failures. A major weakness of combining a large set of potential behaviors with operational scenarios is the inefficiency of a manual evaluation of the multitude of potentially hazardous scenarios generated. Bagschik et al.~\cite{bagschik_identification_2016} found for their use case that several individual scenarios have been created that lead to the same hazards and receive the same risk classification. A main reason for the generation of redundant scenarios is the fact that multiple potential malfunctions lead to the same externally observable malfunctioning behavior of an automated vehicle.

In order to avoid generating redundant scenarios in the context of complex automated driving systems, we propose to adapt the process of identifying the potential malfunctioning behaviors. Since hazards are defined on a vehicle level, the degrees of freedom of the ego-vehicle within operational scenarios are limited to the externally observable vehicle motion. Additionally, every operating scenario includes the description of a desired externally observable behavior, which does not lead to any hazards. Applying this, the proposed hazard identification approach infers malfunctioning behaviors from potential deviations from the desired vehicle behavior. Thus, the systematic generation of potentially hazardous scenarios is not affected by equivalently effective malfunctions of the system.

The potential decrease in size of the dataset of scenarios to be subsequently assessed can be expressed in mathematical terms:
Let $M$ be a set of all malfunctioning behaviors due to individual malfunctions and let $S$ be a set of distinct scenes in operational scenarios, the set of systematically created potentially hazardous scenarios $P_M$ has the size of the Cartesian product $|M\times S|=|P_M|$. With regard to the identification of concrete hazards $f:P_M\rightarrow H$ Bagschik et al.~\cite{bagschik_identification_2016} found that $|P_M|\gg |H|$.

The set of deviations from the desired externally observable vehicle behavior $D$ corresponds with the injection of malfunctioning behavior due to malfunctions $g:M\rightarrow D$. For complex automated driving systems $|M|>|D|$ was observed, with $g(M)\subseteq D$, as some specific deviation might not be caused by an individual malfunction. However, the analysis of the full set of potential deviations is still beneficial in the context of behavioral safety. Applying potential deviations, the set of systematically created potentially hazardous scenarios results from a Cartesian product of scenes and deviations $|P_D|=|D\times S|$. For SAE level 3+ vehicles, we find that $|P_M|>|P_D|$, while $f(P_M )\subseteq f(P_D )$.

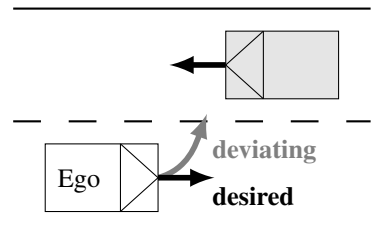
\begin{figure}[thpb]
      \centering
      \usetikzlibrary{arrows, shapes, positioning, shapes.geometric, decorations.markings}
\definecolor{darkgrey}  {RGB} {127,127,127}
\definecolor{lightgrey} {RGB} {229,229,229}
\tikzstyle{lightgreybox} = [draw=black, fill=lightgrey, minimum height=1cm, rectangle, rounded corners, align=center]
\begin{tikzpicture}[x=1mm, y=1mm]

\node (v1) at (0,0) {};
\node (v2) at (50,0) {};
\node (v3) at (0,30) {};
\node (v4) at (50,30) {};
\draw [thick] (v1) -- (v2);
\draw [thick] (v3) -- (v4);
\node (v5) at (0,15){};
\node (v6) at (50,15){};
\draw [thick, dash pattern=on10.5pt](v5) -- (v6);
\draw [-latex,line width=0.75mm,draw,color=darkgrey](20.5,7.5) [out=15, in=250] to (27,16);
\draw [fill=white] (5.5,3) rectangle (20.5,12);
\node (ego) at (10,7) {Ego};
\draw (20.5,7.5) -- (15.5,3);
\draw (15.5,12) -- (15.5,3);
\draw (15.5,12) -- (20.5,7.5);
\draw [-latex,line width=0.75mm,draw] (20.5,7.5) -- (28,7.5);
\node [anchor=west,font=\bfseries](desb) at (26.5,5) {desired};
\node [anchor=west,color=darkgrey,font=\bfseries](devb) at (26.5,11) {deviating};
\draw [fill=lightgrey](44.5,27) rectangle (29.5,18);
\draw (29.5,22.5) -- (34.5,27);
\draw (34.5,18) -- (34.5,27);
\draw (34.5,18) -- (29.5,22.5);
\draw [-latex,line width=0.75mm,draw] (29.5,22.5) -- (22,22.5);
\end{tikzpicture}
      \caption{Operational scenario of oncoming traffic on a two-lane road}
      \label{fig:Scenario}
   \end{figure}

The presence of equivalently effective malfunctions can be illustrated using the scenario depicted in Fig.~\ref{fig:Scenario}. The automated ego-vehicle is following its lane with oncoming traffic in the opposite lane. In order to perform a comprehensive safety conceptualization, the item definition includes all components of the driving functionality (cf.~\cite{winner_towards_2018} for a detailed discussion of item boundaries for a vehicle guidance system). Within the specified operational scenario, both vehicles follow their own lane without interference.
A distinct deviation from the desired behavior of the ego-vehicle is improper and undesired yawing resulting in an entry into the opposite lane.
Recognizably, lane departure results in a hazardous scenario comprising the potential hazard of a crash with the oncoming vehicle. 

By identifying the malfunctioning behaviors of the vehicle based on potential deviations from the specified desired behavior in a scenario, undesired and improper yawing is one potential deviation, independent of specific functionalities.
Thus, a systematic combination of distinct externally observable malfunctioning behaviors with the scenario of Fig.~\ref{fig:Scenario} generates the hazardous scenario of lane departure and collision with oncoming traffic only once.

In contrast, when combining potential malfunctions with operational scenarios, the results can contain one specific external behavior, such as the improper and undesired yawing, multiple times.
In the example presented, the list of potential system malfunctions generating the same potentially hazardous scenario of improper yawing during lane following contains, among others:

\begin{itemize}

\item Inaccurate lane detection
\item Inaccurate map-relative localization
\item Erroneous target pose or trajectory planning
\item Erroneous trajectory tracking
\item Inaccurate ego-motion estimation
\item Erroneous acceleration or brake signals at any wheel
\item Erroneous steering signals
\item Erroneous steering, brake, or drive actuation
\item Defects of wheel or actuation systems

\end{itemize}

To avoid inefficiencies, strategies can be developed to identify redundant hazards before performing an expert-based risk assessment.
However, there are limitations for automatic filtering.
For example, not every potential collision with a specific traffic participant or object is part of the same hazardous scenario and, thus, can be handled as an equivalent hazard.
The externally observable behaviors of the vehicle might differ in specific hazardous scenarios (e.g., collision with an obstacle due to insufficient braking versus no braking).

Our proposed strategy for hazard identification is based on operational scenarios, as defined in the ISO~26262 development process.
In the scenario definition, we include a concrete description of the desired externally observable behavior for the vehicle under development.
Subsequently, we introduce deviating vehicle behaviors in the selected scenarios and analyze the potential consequences.
We do not integrate individual potential malfunctions when generating potentially hazardous scenarios, but use the functional range described in the item definition to analyze the potential externally observable behaviors of the vehicle deviating from its desired functionality.
In principle, deviating vehicle behaviors in a scenario can be any physically possible vehicle motion, considering the intended functional range.
In the context of automated vehicles the major longitudinal and lateral deviations are:

\begin{itemize}

\item Absence of required acceleration
\item Absence of required deceleration
\item Absence of required course angle changes
\item Improper acceleration
\item Improper deceleration
\item Improper course angle changes

\end{itemize}

The potential deviating vehicle behaviors create potentially hazardous scenarios in the context of the operational scenarios.
Subsequently, the potentially hazardous scenarios are checked in an expert-based evaluation.
In order to form hazardous scenarios, the presence of all three components of a hazard has to be verified: source, target, and initiating mechanism~\cite{ericson_hazard_2005}.
The analyzed set of hazardous scenarios is free of redundant hazards, as the scenarios are based on different externally observable behaviors of the ego-vehicle.
Based on the scenarios, we identify and describe hazards for subsequent risk assessment.
During hazard identification, several consequences of each potentially hazardous scenario can be analyzed and documented, creating one or many hazardous scenarios and hazards (e.g., a hazard of a collision with infrastructure and a hazard of a collision with a pedestrian).

\section{CASE STUDY: HAZARD IDENTIFICATION IN UNICARAGIL}

\label{sec:usecase}

The presented hazard identification strategy is developed as part of the safety concept generation for the automated driving function in the research project UNICAR\emph{agil}. 
In UNICAR\emph{agil}, funded by the German Federal Ministry of Education and Research (BMBF), a project consortium of institutes from eight German universities is aiming to showcase four driverless cars for inner city traffic by 2022.
Six industrial partners support the development.
The four UNICAR\emph{agil} vehicles are based on a common modular platform, but are designed for diverse use cases (cf.~\cite{woopen_unicaragil_2018} for a detailed concept and use case description).

A common driving function is developed and used for the UNICAR\emph{agil} vehicles, as they share the technological platform and design domain.
Therefore, the developed safety concept for the driving function applies to all of the four realized vehicle types.
In an early design stage, we mainly aim to analyze the abstract behavioral safety aspects and not specific functional safety concerns of individual components.
In this regard, Klamann et al.~\cite{klamann_defining_2019} detail the different levels of abstraction for microscopic safety requirements employed within UNICAR\emph{agil}.

On the top level, we express a desired externally observable vehicle behavior in abstract safety goals that form the foundation of the safety concept.
So far, we have systematically identified hazards in representative traffic scenarios and subsequently performed risk assessments in order to define a first set of necessary safety goals.
In this section we describe one of the scenarios used in the concept phase of development in UNICAR\emph{agil} and show the results of an application of our proposed hazard identification strategy.

   \begin{figure*}[thpb]
      \centering
      \usetikzlibrary{arrows, shapes, positioning, shapes.geometric, decorations.markings, patterns}
\definecolor{darkgrey}  {RGB} {127,127,127}
\definecolor{lightgrey} {RGB} {229,229,229}
\tikzstyle{lightgreybox} = [draw=black, fill=lightgrey, minimum height=1cm, rectangle, rounded corners, align=center]
\begin{tikzpicture}[x=1mm, y=1mm]

\node (v1) at (0,0) {};
\node (v2) at (100,0) {};
\node (v3) at (0,30) {};
\node (v4) at (100,30) {};
\draw [thick] (v1) -- (v2);
\draw [thick] (v3) -- (v4);
\node (v5) at (0,15){};
\node (v6) at (100,15){};
\draw [thick, dash pattern=on10.5pt](v5) -- (v6);
\draw (7.5,3) rectangle (22.5,12);
\node (ego) at (12,7) {Ego};
\draw (22.5,7.5) -- (17.5,3);
\draw (17.5,12) -- (17.5,3);
\draw (17.5,12) -- (22.5,7.5);
\draw [line width=0.3mm,draw,dash pattern=on4pt] (27.5,7.5)[in=180,out=0] to (50,5) to (60,5);
\draw [-latex,line width=0.75mm,draw] (22.5,7.5) -- (30,7.5);
\node (v0) at (30,10) {$v_{\mathrm{Ego},0}$};
\draw  [dash pattern=on3pt](60,0.5) rectangle (75,9.5);
\node (ego) at (12,7) {Ego};
\draw [dash pattern=on3pt](75,5) -- (70,0.5);
\draw [dash pattern=on3pt](70,9.5) -- (70,0.5);
\draw [dash pattern=on3pt](70,9.5) -- (75,5);
\draw [-latex,line width=0.75mm,draw] (75,5) -- (81,5);
\node (vpass) at (83,7.5) {$v_{\mathrm{Ego},\mathrm{pass}}$};

\draw [pattern=north west lines, pattern color=black] (42.5,39) rectangle (27.5,30.5);
\draw [pattern=north west lines, pattern color=black] (62.5,39) rectangle (47.5,30.5);
\draw [pattern=north west lines, pattern color=black] (95,39) rectangle (70,30.5);
\draw [line width=0.2mm,draw,dash pattern=on2.5pt] (66.5,33) -- (66.5,20);
\draw [-latex,line width=0.5mm,draw] (66.5,33) -- (66.5,26);
\node (vpass) at (72,26) {$v_{\mathrm{P},0}$};
\draw (66.5,33)[fill=lightgrey] circle (2.5);
\node (Ped) at (66.5,33) {P};
\draw (66.5,17.35)[dash pattern=on2pt] circle (2.5);
\end{tikzpicture}
      \caption{Illustration of an operational reference scenario in UNICAR\emph{agil}. Pedestrian~P enters the road from an occlusion and halts for the automated ego-vehicle to pass. The ego-vehicle passes the waiting pedestrian with adjusted speed and lateral position.}
      \label{fig:Scenario_UNICAR}
   \end{figure*}
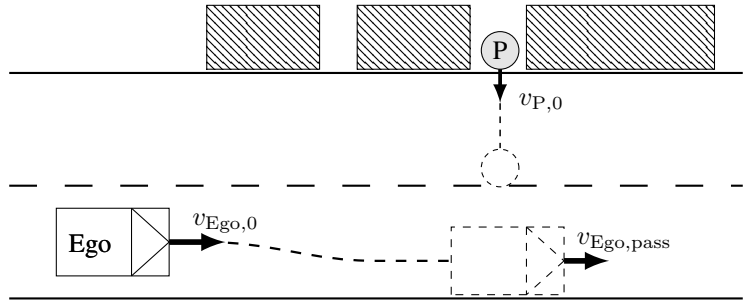

\subsection{Scenario Description}
The analyzed scenario ``Occluded Pedestrian'' is depicted in Fig.~\ref{fig:Scenario_UNICAR}:
the ego-vehicle travels with the speed $v_{\mathrm{Ego},0}$ in the center of its own lane on a two-lane inner city road.
A pedestrian~P enters the neighbor lane from an occlusion and moves towards the center lane marking with $v_{\mathrm{Ped},0}$. 
To account for the uncertainty in the predicted movement and intentions of other traffic participants, the automated ego-vehicle reduces its speed (cf.~\cite{bagschik_systems_2018} for a deeper discussion of risk-aware selection of an adequate travel speed).

At a later time step, the pedestrian halts at the center lane marking for the ego-vehicle to pass.
To account for the uncertainties in the relative position and future movement of the pedestrian, the ego-vehicle maximizes the lateral offset within its own lane and passes at the low speed $v_{\mathrm{Ego},\mathrm{pass}}$.

The desired externally observable behavior of the ego-vehicle in the presented operational scenario corresponds with several of the defined competencies that are required from automated vehicles:
using the list of behavioral competencies of the U.S. Department of Transportation's Federal Automated Vehicles Policy~\cite{national_highway_traffic_safety_administration._federal_2016}, ``Provide Safe Distance From Vehicles, Pedestrians, Bicyclists on Side of the Road'' is one of the competencies the automated vehicle demonstrates in our scenario.
The scenario moreover corresponds to ``Detect and Respond to Pedestrians in Road (Not Walking Through Intersection or Crosswalk),'' which is one of the additional behavioral competencies listed in the Waymo Safety Report~\cite{waymo_waymo_2017}.

\subsection{Deviating Vehicle Behaviors}
The subsequent step of our proposed hazard identification strategy is the introduction of deviating vehicle behaviors in the analyzed scenario.
Here, the special dynamic characteristics of the UNICAR\emph{agil} vehicles had to be considered: wheel individual all-wheel steering with wheel angles up to 90~degrees allows sudden alterations of the course angle without significant path curvature or yawing.
To form potentially hazardous scenarios, we used the generic list of deviations from Section~\ref{sec:proposedstrat} to find applicable behavior deviations of the ego-vehicle in the scenario ``Occluded Pedestrian'': 
\begin{itemize}

\item Absence of required speed adjustment
\item Absence of required lateral position adjustment
\item Improper acceleration at any moment
\item Improper (rapid) deceleration
\item Improper course angle changes

\end{itemize}

\subsection{Hazards}
To finally specify all involved hazards, the deviating behaviors in the context of the described scenario (i.e.~potentially hazardous scenarios) have to be evaluated by experts.
In UNICAR\emph{agil} the potentially hazardous scenarios were analyzed for actual hazards to health and life of vehicle passengers and other traffic participants.
In other scenarios, clear violations of traffic law were also included in the list of identified hazards  (e.g., ignoring stop signs is considered general hazardous behavior).

Hazards to health and life of the pedestrian were found in the following hazardous scenarios:
\begin{itemize}

\item The vehicle does not adjust its speed when approaching the pedestrian. The combination of inadequate traveling speed and present uncertainties leads to a collision between vehicle and pedestrian.
\item The vehicle accelerates when approaching the pedestrian. The combination of high traveling speed and present uncertainties leads to a collision between vehicle and pedestrian.
\item The vehicle changes its course angle towards the pedestrian, leading to a collision between vehicle and pedestrian.

\end{itemize}

Hazards to health and life of vehicle passengers were found in the following hazardous scenarios:
\begin{itemize}

\item The vehicle decelerates rapidly during lane following.
\item The vehicle changes its course angle towards the roadside, leading to a collision between vehicle and stationary infrastructure.

\end{itemize}

\section{COMPARISON OF HAZARD IDENTIFICATION STRATEGIES}

\label{sec:comparison}

\begin{figure*}[thpb]
      \centering
      \subfloat[Systematic hazard identification approach applied by Bagschik et al.~\cite{bagschik_identification_2016}\label{fig:ISOProcessBagschik}]{\usetikzlibrary{arrows, shapes, positioning, shapes.geometric, decorations.markings}
\definecolor{darkgrey}  {RGB} {127,127,127}
\definecolor{lightgrey} {RGB} {229,229,229}
\tikzstyle{lightgreybox} = [draw=black, fill=lightgrey, minimum height=1cm, rectangle, rounded corners, align=center]
\begin{tikzpicture}[x=1mm, y=1mm]

\node [minimum height=0.5cm, rounded corners, minimum width = 5cm, text centered, font=\bfseries\large] (idef) {Item definition};
\node [lightgreybox, minimum height=1cm, minimum width = 2.375cm,text depth=0.25ex,  below left=0.75cm and -2.375cm of idef] (sce) {Operational \\scenarios};
\node [lightgreybox, minimum height=1cm, minimum width = 2.375cm, text depth=0.25ex,  below= 0.5cm of sce] (ssce) {Operational \\scenes};
\draw [-latex,thick,draw] (sce) -- (ssce) node [above, midway, anchor = south] {};
\node [lightgreybox, minimum height=1cm, minimum width = 2.375cm, text height=1.5ex, text depth=0.25ex, below  right=0.75cm and-2.375cm of idef] (func) {Functions};
\draw [-latex,line width=0.75mm,draw] (idef.south-|sce) -- (sce) node [above, midway, anchor = south] {};
\draw [-latex,line width=0.75mm,draw] (idef.south-|func) -- (func) node [above, midway, anchor = south] {};
\node [lightgreybox, minimum height=1cm, minimum width = 2.375cm, text height=1.5ex, text depth=0.25ex, below=0.5cm of func] (malf) {Malfunctions};
\draw [-latex,thick,draw] (func) -- (malf) node [above, midway, anchor = south] {};
\node [lightgreybox, fill=white, line width=2pt, minimum height=1.25cm, minimum width = 5cm,text depth=0.5ex, below left=0.5cm and -2.375cm of malf] (even) {Systematic generation of\\ potentially hazardous scenarios};
\draw [-latex,thick,draw] (ssce) -- (even.north-|ssce) node [above, midway, anchor = south] {};
\draw [-latex,thick,draw] (malf) -- (even.north-|malf) node [above, midway, anchor = south] {};
\node [lightgreybox,  fill=white, minimum height=1.25cm, minimum width = 5cm, text depth=0.5ex, below=0.5cm of even] (eval) {Expert-based evaluation\\ of hazardous scenarios};
\draw [-latex,thick,draw] (even) -- (eval) node [above, midway, anchor = south] {};
\node [lightgreybox, minimum height=1cm, minimum width = 5cm, text height=1.5ex, text depth=0.25ex, below=0.5cm of eval] (haza) {Identified hazards};
\draw [-latex,thick,draw] (eval) -- (haza) node [above, midway, anchor = south] {};
\node [minimum height=0.5cm, rounded corners, minimum width = 5cm, below=0.75cm of haza, text centered, font=\bfseries\large] (rass) {Risk assessment};
\draw [-latex,line width=0.75mm,draw] (haza) -- (rass) node [above, midway, anchor = south] {};

\end{tikzpicture}} \hspace{0.235cm}
      \subfloat[Expert-based hazard identification performed in~\cite{stolte_hazard_2017}\label{fig:ISOProcessExpert}]{\usetikzlibrary{arrows, shapes, positioning, shapes.geometric, decorations.markings}
\definecolor{darkgrey}  {RGB} {127,127,127}
\definecolor{lightgrey} {RGB} {229,229,229}
\tikzstyle{lightgreybox} = [draw=black, fill=lightgrey, minimum height=1cm, rectangle, rounded corners, align=center]
\begin{tikzpicture}[x=1mm, y=1mm]

\node [minimum height=0.5cm, rounded corners, minimum width = 5cm, text centered, font=\bfseries\large] (idef) {Item definition};
\node [lightgreybox, minimum height=2.5cm, minimum width = 2.375cm, below left=0.75cm and -2.375cm of idef] (sce) {Operational \\design \\domain};
\node [lightgreybox, minimum height=1cm, minimum width = 2.375cm, text height=1.5ex, text depth=0.25ex, above right=-1cm and 0.25cm of sce] (func) {Functions};
\draw [-latex,line width=0.75mm,draw] (idef.south-|sce) -- (sce) node [above, midway, anchor = south] {};
\draw [-latex,line width=0.75mm,draw] (idef.south-|func) -- (func) node [above, midway, anchor = south] {};
\node [lightgreybox, minimum height=1cm, minimum width = 2.375cm, text height=1.5ex, text depth=0.25ex, below right=-1cm and 0.25cm of sce] (malf) {Malfunctions};
\draw [-latex,thick,draw] (func) -- (malf) node [above, midway, anchor = south] {};
\node [lightgreybox, fill=white, line width=2pt, minimum height=3cm+1pt, minimum width = 5cm,text depth=0.5ex, below right=0.5cm and -2.375cm of sce] (even) {Expert-based identification\\ and evaluation of\\ hazardous scenarios};
\draw [-latex,thick,draw] (sce) -- (even.north-|sce) node [above, midway, anchor = south] {};
\draw [-latex,thick,draw] (malf) -- (even.north-|malf) node [above, midway, anchor = south] {};
\node [lightgreybox, minimum height=1cm, minimum width = 5cm, text height=1.5ex, text depth=0.25ex, below=0.5cm of even] (haza) {Identified hazards};
\draw [-latex,thick,draw] (even) -- (haza) node [above, midway, anchor = south] {};
\node [minimum height=0.5cm, rounded corners, minimum width = 5cm, below=0.75cm of haza, text centered, font=\bfseries\large] (rass) {Risk assessment};
\draw [-latex,line width=0.75mm,draw] (haza) -- (rass) node [above, midway, anchor = south] {};

\end{tikzpicture}} \hspace{0.235cm}
      \subfloat[Proposed hazard identification approach\label{fig:ExtBehProcess}]{\usetikzlibrary{arrows, shapes, positioning, shapes.geometric, decorations.markings}
\definecolor{darkgrey}  {RGB} {127,127,127}
\definecolor{lightgrey} {RGB} {229,229,229}
\tikzstyle{lightgreybox} = [draw=black, fill=lightgrey, minimum height=1cm, rectangle, rounded corners, align=center]
\begin{tikzpicture}[x=1mm, y=1mm]

\node [minimum height=0.5cm, rounded corners, minimum width = 5cm, text centered, font=\bfseries\large] (idef) {Item definition};
\node [lightgreybox, minimum height=1cm, minimum width = 5cm,text depth=0.25ex,   below =0.75cm of idef] (sce) {Operational scenarios\\ with desired vehicle behavior};
\draw [-latex,line width=0.75mm,draw] (idef.south-|sce) -- (sce) node [above, midway, anchor = south] {};
\node [lightgreybox, minimum height=1cm, minimum width = 5 cm, text height=1.5ex, text depth=0.25ex, below =0.5cm of sce] (devb) {Deviating vehicle behaviors};
\draw [-latex,thick,draw] (sce) -- (devb) node [above, midway, anchor = south] {};
\node [lightgreybox, fill=white, line width=2pt, minimum height=1.25cm, minimum width = 5cm,text depth=0.5ex, below =0.5cm of devb] (even) {Systematic generation of\\ potentially hazardous scenarios};
\draw [-latex,thick,draw] (devb) -- (even) node [above, midway, anchor = south] {};
\node [lightgreybox,  fill=white, minimum height=1.25cm, minimum width = 5cm, text depth=0.5ex, below=0.5cm of even] (eval) {Expert-based evaluation\\ of hazardous scenarios};
\draw [-latex,thick,draw] (even) -- (eval) node [above, midway, anchor = south] {};
\node [lightgreybox, minimum height=1cm, minimum width = 5cm, text height=1.5ex, text depth=0.25ex, below=0.5cm of eval] (haza) {Identified hazards};
\draw [-latex,thick,draw] (eval) -- (haza) node [above, midway, anchor = south] {};
\node [minimum height=0.5cm, rounded corners, minimum width = 5cm, below=0.75cm of haza, text centered, font=\bfseries\large] (rass) {Risk assessment};
\draw [-latex,line width=0.75mm,draw] (haza) -- (rass) node [above, midway, anchor = south] {};

\end{tikzpicture}}
      \caption{Stages of the discussed strategies for hazard identification. White boxes indicate the actual process steps for hazard identification, those with bold outlines mark the source of hazardous scenarios.}
      \label{fig:Comparison}
   \end{figure*}
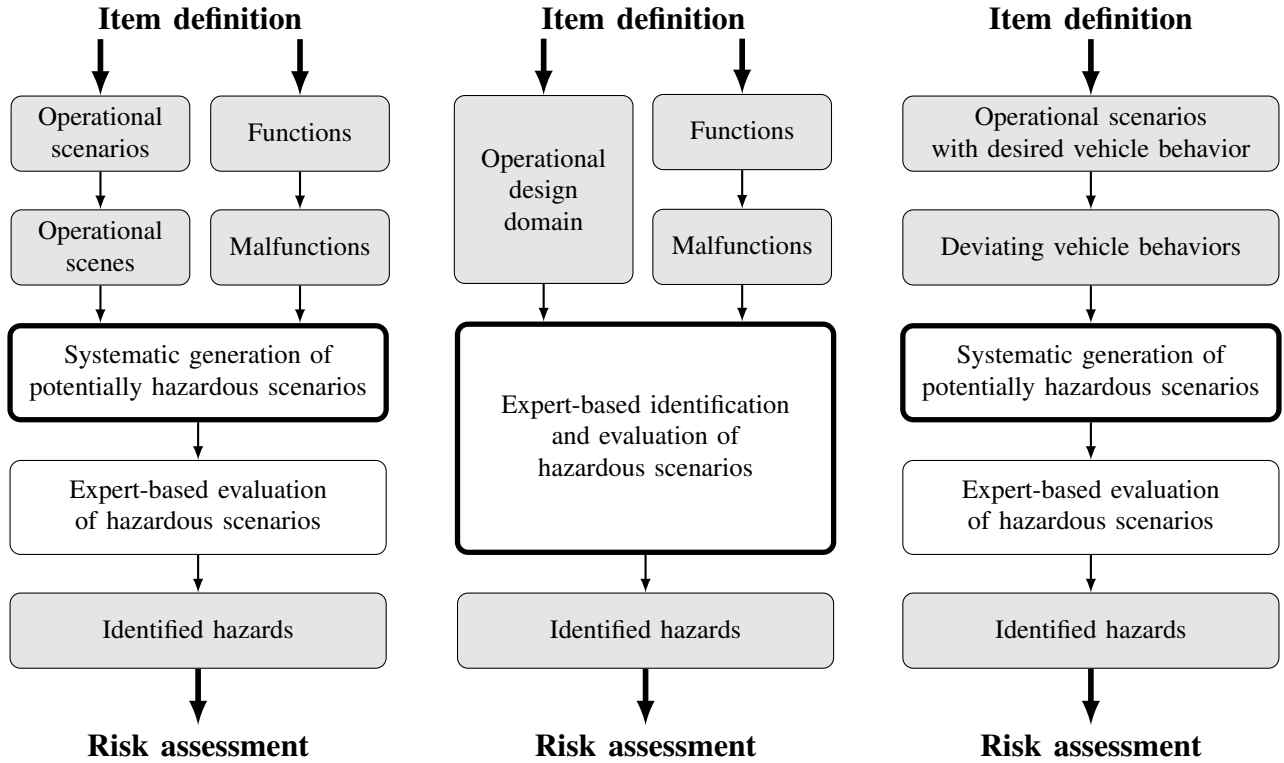

In order to compare the proposed strategy with other published approaches, Fig.~\ref{fig:Comparison} depicts the individual stages of three different strategies for hazard identification.
As we are acutely aware of the course of the research project of an unmanned protective vehicle for hard shoulder roadworks and can profoundly highlight differences to our ongoing research, we again use publications about this project as a reference for other approaches.
Many of the publications mentioned in Section~\ref{ssec:relatedwork}, however, also represent one of the two generic strategies illustrated in Fig.~\ref{fig:ISOProcessBagschik} (i.e.~\cite{pegasus_project_office_pegasus_nodate,lurie_hazard_2018,beckers_structured_2013,becker_functional_2018}) and Fig.~\ref{fig:ISOProcessExpert} (i.e.~\cite{valdez_banda_systemic_2019,aceituna_prehaz:_2019,sexton_effective_2014}).

The different processes share the input of information regarding the functional range of the system under development from an item definition.
Also, all three approaches eventually yield a set of identified hazards for subsequent risk assessment.
In the following paragraphs, the process flows of the individual strategies are discussed in detail.

\subsection{Combination of Malfunctions and Operational Scenes}

The general process flow of the hazard identification approach presented by Bagschik et al.~\cite{bagschik_identification_2016} is depicted in Fig.~\ref{fig:ISOProcessBagschik}.
Following the development process of ISO~26262, the item definition initially describes functions and operational scenarios to be analyzed for hazard identification.
As a next process step, they infer potential malfunctions for each functionality of the item.
Known malfunctions or failure modes of a system under development can already be described in the item definition and support this task.

In addition, Bagschik et al.~\cite{bagschik_identification_2016} define operational scenes by splitting potential operational scenarios into multiple distinct scenes.
Subsequently, possible malfunctions are put in the context of the individual scenes to create potentially hazardous scenarios.
The set of scenarios is analyzed manually for sources of potential harm in order to identify hazardous scenarios.
An expert-based evaluation of the hazardous scenarios detects and documents potential hazards for further examination in the context of a risk assessment.

Bagschik et al.~\cite{bagschik_identification_2016} apply their strategy to identify hazards for the operation of an unmanned protective vehicle for hard shoulder roadworks.
The limited functional range of the automated driving function of their use case still results in 37 identified potential malfunctions, 108 relevant scenes, and more than 21,000 potentially hazardous scenarios.
After automatic filtering the dataset contains 750 scenarios that need to be manually assessed to evaluate and describe the included hazards.

\subsection{Expert-based Hazard Identification}

The process flow of an expert-based hazard identification based on Stolte et al.~\cite{stolte_hazard_2017} is depicted in Fig.~\ref{fig:ISOProcessExpert}.
Again, potential system malfunctioning behaviors are derived from planned functions, which are described in the item definition. 

In contrast to other strategies, Stolte et al. do not explicitly define operational scenarios before performing the hazard identification.
The expert-identification of hazardous scenarios is based on the operational design domain described in the item definition.
Following SAE~J3016~\cite{sae_international_sae_2018}, the operational design domain defines the operating conditions under which an automated vehicle is designed to function, including environmental and geographical restrictions, as well as traffic and roadway characteristics.

Based on the operational design domain and the potential malfunctioning behaviors, experts identify hazardous scenarios and evaluate potential consequences in one process step of the hazard analysis and risk assessment process presented by Stolte et al.
Analyzing the results of this task, 14 distinct operational scenarios can be identified that were starting points of different hazardous scenarios.
The considered operational scenarios differ in parameters such as system status, ego-vehicle position and velocity, hard shoulder setup, traffic situation, etc.

\subsection{Proposed Hazard Identification Approach}

The process flow of our proposed strategy described in Section~\ref{sec:proposedstrat} is displayed in Fig.~\ref{fig:ExtBehProcess}. 
In contrast to the other strategies, system malfunctions are not explicitly described or used for hazard identification.
Essentially, we introduce generic longitudinal and lateral deviations in operational scenarios and evaluate their consequences.

The use of deviations from a demanded safe behavior for hazard identification has also been described by Bagschik et al.~\cite{bagschik_systems_2018}.
They analyze the externally observable lateral and longitudinal behavior of an automated vehicle when approaching a pedestrian crossing.
In contrast to the process depicted in Fig.~\ref{fig:ExtBehProcess}, Bagschik et al.~\cite{bagschik_systems_2018} first specify an accident within the operational scenario of striking and fatally injuring a crossing pedestrian.
Subsequently, they derive hazards directly from potential deviations of vehicle behavior that can cause the specified accident.
On the one hand, their strategy saves the effort of evaluating systematically generated potentially hazardous scenarios.
On the other hand, potential accidents have to be identified and selected for further analysis before potential deviating vehicle behavior as initiating mechanisms of individual hazards are examined. 
Ultimately, the systematic identification of potential accidents within our proposed approach (cf.~Fig.~\ref{fig:ExtBehProcess}) supports the traceability of developed safety requirements, especially when considering different scenarios with several traffic participants.

\subsection{Discussion}

The expert-based hazard identification presented by Stolte et al.~\cite{stolte_hazard_2017} was successfully applied in the context of a very limited operational design domain. Other vehicle automation functionalities will offer fewer restrictions regarding the functional range, which far increases the complexity of a manual identification of potential hazardous scenarios. In addition, the already high number of potential malfunctions of an automated vehicle presumably increases with its functional range. Thus, we argue that the development of a sound set of safety goals for complex automated driving systems ultimately requires a systematic combination of operational scenarios with potential malfunctioning behaviors.

All of the 37 malfunctioning behaviors identified by Bagschik et al.~\cite{bagschik_identification_2016} can be described by one of the deviations from the desired behavior used in the proposed approach (cf.~Section~\ref{sec:proposedstrat}). This reduces the amount of malfunctioning behaviors to be individually combined with operational scenarios to six. Furthermore, filtering can be performed before manually examining the relevant scenarios in an expert-based evaluation of hazardous scenarios. For example, absence of required acceleration, deceleration, and course angle changes is only applicable in specific segments of a scenario.

Altogether, the proposed approach for hazard identification avoids the inefficiency of evaluating identical hazardous scenarios multiple times in an expert-based hazard analysis process. However, the large amount of potentially hazardous scenarios when applying systematic strategies is a product of both, the quantity of operational scenarios and the quantity of potential malfunctions. The number of distinct operational scenarios to consider results from the definition of the operational design domain of the system under development. Current research shows that, even when limiting the operational design domain to one specific road type, the set of operational scenarios created by a systematic generation process is too large to be manually analyzed~\cite{bagschik_wissensbasierte_2018}.

In an early design phase, corresponding with best practice, a reduced set of operational scenarios is analyzed within the hazard analysis and risk assessment task. For this, a systematic generation of reference scenarios for requirement analysis and safety assessment as described by Ebner~\cite{ebner_referenzszenarien_2014} can be integrated. The analysis of a few important scenarios makes it possible to create an initial set of safety goals and to start developing mitigation strategies for the identified hazards. Early first iterations of the concept phase tasks of safety conceptualization are a prerequisite for aligning the functional range described in the item definition with feasible safety requirements~\cite{graubohm_systematic_2017}.

Within the phase of preliminary design, the proposed approach far improves the efficiency of the expert-based hazard evaluation and risk assessment.
Still, the manual examination of potentially hazardous scenarios entails the disadvantage that relevant hazards could be misjudged or omitted.
Within later stages of development, an exhaustive set of scenarios has to be analyzed. Corner cases have to be identified, in which hazardous consequences of malfunctions occur that have not been addressed in the preliminary safety concept.

\subsection{Subsequent Process Steps}

Concluding the hazard analysis and risk assessment task, the identified hazards are classified and safety goals are defined to address the hazards. The next work product of the safety life cycle of ISO~26262 is the generation of a functional safety concept. A distinct feature of functional safety concepts is the consideration of specific functional components when developing mitigation strategies and defining safety requirements on the basis of safety goals.

The hazard identification strategy outlined in ISO~26262 and performed in related work (cf.~Section~\ref{sec:strategies}) simultaneously generates traceable links between hazards and functions: hazardous scenarios are derived from malfunctions, which are associated with specific functionalities. With regard to the results of our proposed hazard identification approach, these links are missing and have to be created a posteriori. Therefore, the deviations of relevant functions (i.e.~malfunctions) that lead to the undesired externally observable behavior in operational scenarios have to be identified.

\addtolength{\textheight}{-0.84349cm}   

\section{CONCLUSION}

\label{sec:conclusion}

In this contribution, we discussed an adaptation of the strategy for hazard identification for driverless vehicle development.
Our main goal was to reduce the amount of potentially hazardous scenarios to be analyzed in an early design phase.
Instead of focusing on malfunctions, we based our approach on a combination of operational scenarios with potential deviations from the desired vehicle behavior.
To indicate the achieved efficiency, we presented the process and results of an application of our identification approach in the ongoing research project UNICAR\emph{agil} as a case study.

In the future, we plan to associate the discussion of a systematic specification of operational scenarios which are of value for early hazard analyses with our presented hazard identification strategy.
The use of a structured process for finding and analyzing the right scenarios will improve the description of a systematic preliminary development of automated vehicles.
Additionally, the presented approach will be further evaluated in the context of the development of specific safety requirements for the UNICAR\emph{agil} vehicles.
One crucial subsequent process step is the systematic derivation of component-related safety requirements and test cases from vehicle level safety goals (cf.~\cite{klamann_defining_2019}).




\section*{ACKNOWLEDGMENT}
We thank Felix Gr\"un for stimulating discussions of the presented work and Sonja Luther for proofreading.

\end{document}